 \documentclass[manuscript]{aastex}

\slugcomment{Submitted to Astronomical Journal}

\shorttitle{Blue Variables from the MACHO database}
\shortauthors{Keller et al.}

\begin{document}


\title{Blue Variable Stars from the MACHO database I: \\ 
Photometry and Spectroscopy of the LMC sample}


\author{S.\ C.\ Keller}
\affil{Institute of Geophysics and Planetary Physics, LLNL,\\
Livermore, CA 94550}
\email{skeller@igpp.ucllnl.org}

\author{M.\ S.\ Bessell}
\affil{Research School of Astronomy and Astrophysics, ANU,\\
Canberra, Australia}

\author{K.\ H.\ Cook}
\affil{Institute of Geophysics and Planetary Physics, LLNL,\\
Livermore, CA 94550}

\author{M.\ Geha}
\affil{Dept. of Astronomy and Astrophysics,\\
UCO/Lick Observatory, Univ. of California, Santa Cruz, CA  95064}

\and

\author{D.\ Syphers}
\affil{Center for Cosmological Physics, University of Chiago,\\ 
5640 S. Ellis Ave., Chicago, IL 60637 }


\begin{abstract}
  We present the photometric properties of 1279 blue variable stars within
  the LMC. Photometry is derived from the MACHO database. The lightcurves of
  the sample exhibit a variety of quasi-periodic and aperiodic outburst
  behavior. A characteristic feature of the photometric variation is that
  the objects are reddest when at maximum outburst. A subset of 102 objects
  were examined spectroscopically. Within this subset, 91\% exhibited Balmer
  emission in at least one epoch, in some cases with spectacular spectral
  variability. The variability observed in the sample is consistent with the
  establishment and maintenance of the Be phenomenon.
\end{abstract}


\section{Introduction}

The MACHO photometric database constitutes an unprecedented resource for the
study of stellar variability. With coverage of over 20 million stars within
the LMC alone, the MACHO project is superior to competitors in duration
(seven years) and spatial coverage. A primary goal of the MACHO project was
to detect microlensing events in/towards the Magellanic Clouds and the bulge
of the Galaxy. One of the initial unexpected contaminants in the detection
of microlensing events were blue variables provisionally termed ``bumpers''
\citep{coo95}. These objects were found to brighten $\Delta V$ $\sim$
0.$^m$2-0.4 over the course of fifty to several hundred days in a manner
much like that expected from microlensing events. However these objects did
not exhibit the achromaticity expected of microlensing.

%Further consideration revealed that the spectra of many bumpers showed
%Balmer emission i.e.\ the majority of bumpers are Be stars.
In the present work we extract from the catalogue of MACHO variables (over
210000 objects) those stars which are in the vicinity of the main-sequence
and exhibit large amplitude variability (see Section \ref{sect:sel}).
Section \ref{sect:phot} presents an overview of lightcurve morphology and
colour variations. Section \ref{sect:spec} presents spectroscopic
observations of a subset of 102 objects from our catalogue. In Section
\ref{sect:nature} we discuss the physical nature of the blue variables. The
goal of the current paper is to present preliminary data on these objects
for future studies to expand upon.

\section{Selection of the Sample}
\label{sect:sel}
Instrumental MACHO $B$ and $R$ magnitudes have been transformed to standard
Johnson $V$ and Cousins $R$ (for details see Alcock et al.\ 1999).
From this database we have selected blue (i.e.  -0.5$<V$$-$$R<$0.4) objects
with $V<18.0$.  We then select those stars with significant variability.
Our variability index borrows from that used to detect MACHO microlensing
events \citep{alc95}.  Under this scheme, a star is flagged as
variable if the lightcurve contains at least 10 pairs of $V$ and $R$ of
large $\chi^{2}$ compared to a constant $V$ and $R$ brightness. From the 30
LMC MACHO fields with calibrated photometry this results in a sample of 2841
candidate variables.  A major contaminant are eclipsing binaries which were
excluded by visual inspection. This results in a sample of 1279 objects. The
positions, magnitudes and colours of these objects are presented in Table
\ref{tab:data} (full table is available from ADS).

\section{Photometry}
\label{sect:phot}
Our selected blue variables show a remarkable variety of lightcurve
morphology. To attempt to classify each star into a series of arbitrary
categories is perhaps misleading as stars may show combinations of what
might be termed ``modes'' of variability. These ``modes'' can be described as
follows:

1. Bumper events - these events are typically $\Delta V$=0.$^m$2-0.$^m$4. in
amplitude and duration of 100-800 days.

2. Flicker events - rapid, generally low amplitude ($\Delta
V$=0.$^m$05-0.$^m$15) variation. Duration of outburst ranges from 10-50
days. Rise time is very short (several days) followed by ``exponential''
decay. Larger outbursts result in longer durations. Flickering behavior is
often associated with maximum light of bumper modes (Fig.\ \ref{figpage2}).

3. Step events - within the course of 10-50 days the brightness increases
$\Delta V$=0.$^m$2-0.$^m$3.

4. Baseline variation - long-term trend of 1000s of days with amplitudes
up to $\Delta V\sim$0.$^m$4.

5. Fading events - drop in $V$ luminosity of up to $\Delta V\sim$0.$^m$4 with
time scales 200-600 days.

\begin{figure}
\plotone{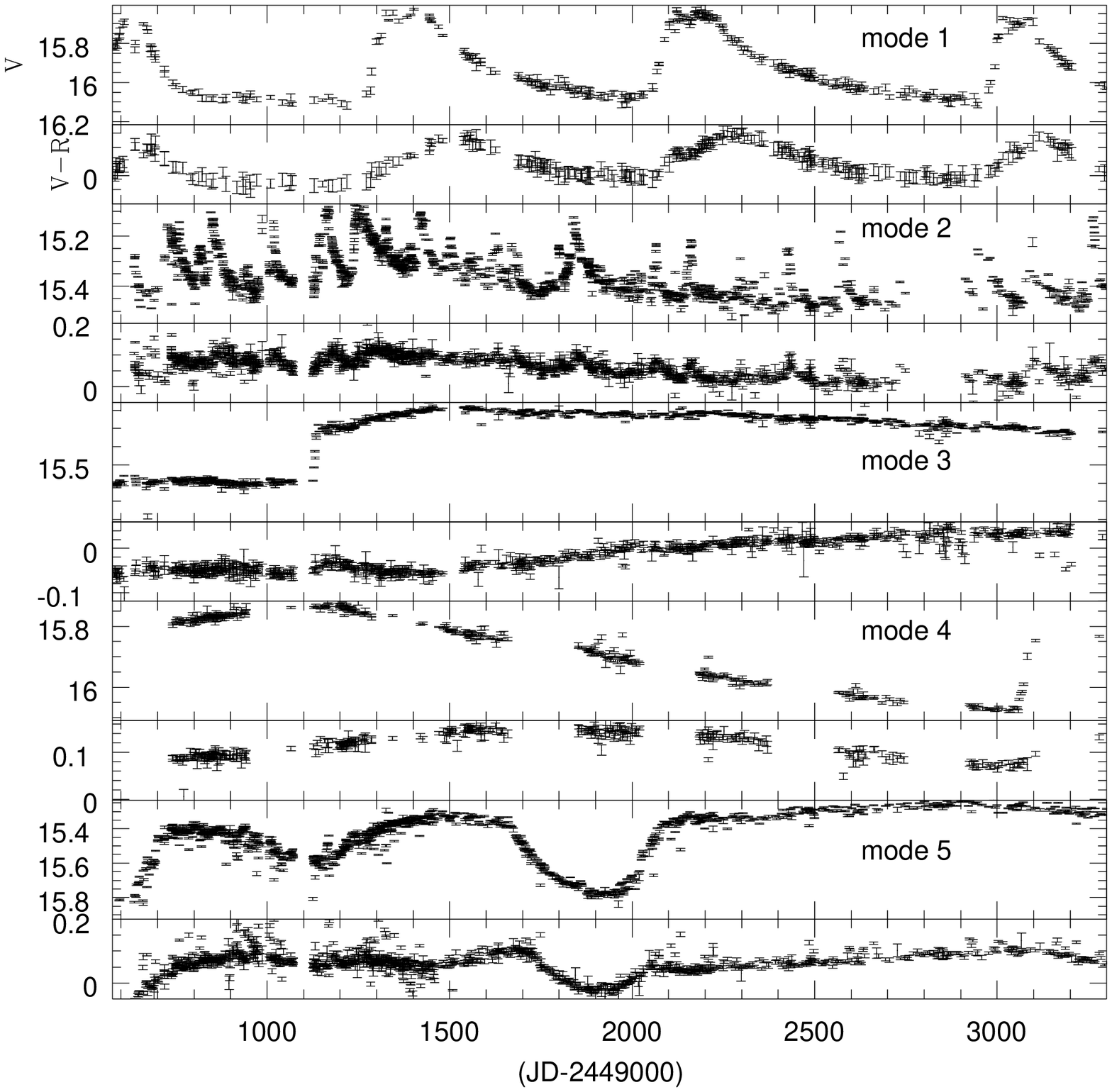}
\caption{$V$ lightcurves and $V$$-$$R$ colourcurves for examples of the
  five modes of variability exhibited by the sample of blue variables. The
  MACHO identifications from top to bottom are 17.2109.68, 77.7672.34,
  14.8500.382, 81.9490.41, and 1.4050.1666.\label{figpage1}}
\end{figure}

\begin{figure}
\plotone{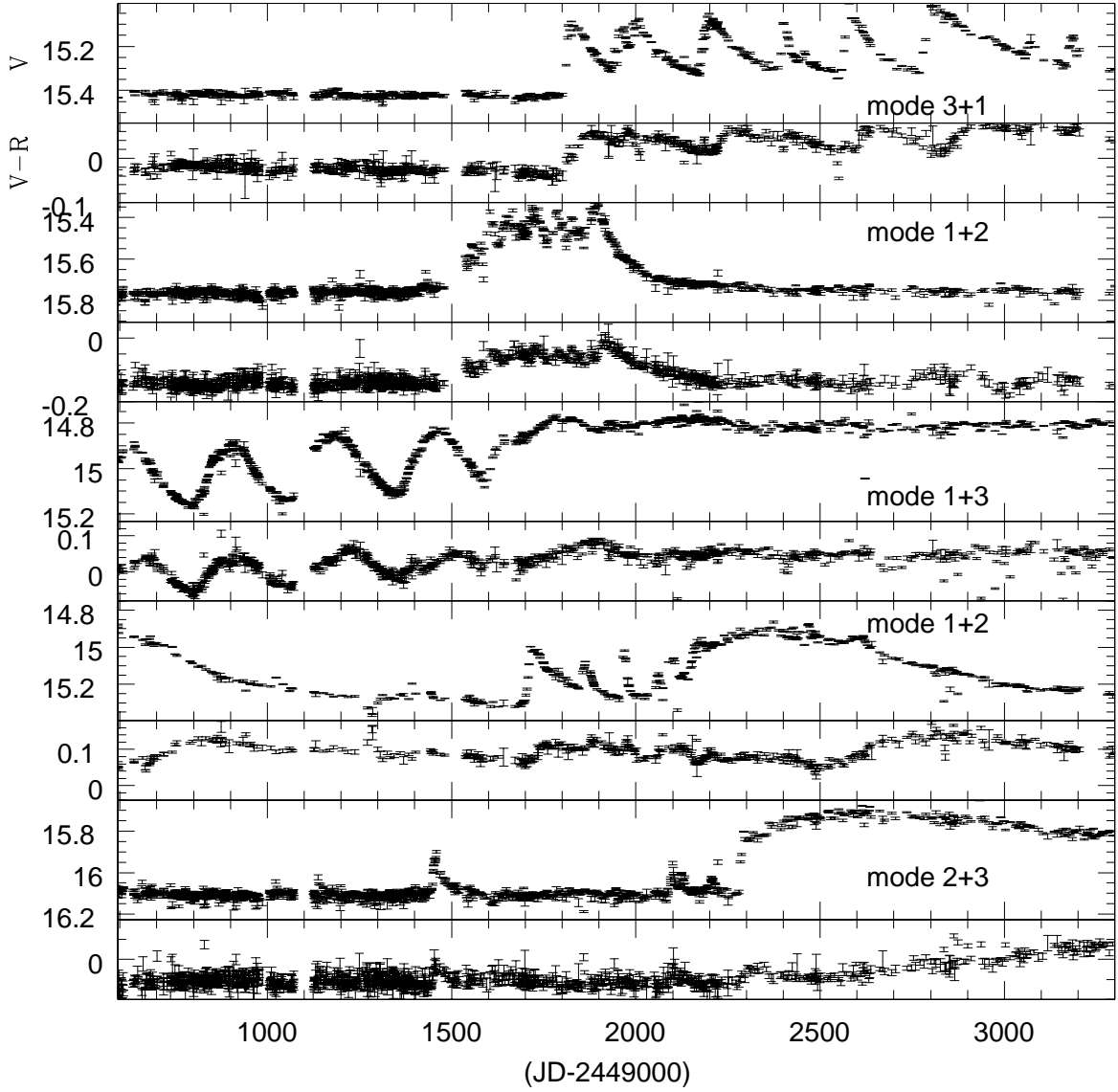}
\caption{Example $V$ lightcurves and $V$$-$$R$ colourcurves for objects
  showing the variety of behaviour obtained from juxtaposition of the five
  modes of variability defined in Fig.\ \ref{figpage1}. The MACHO
  identifications from top to bottom are 10.3195.13, 1.3813.48, 10.4036.11, 17.2229.18, and 1.3931.95 \label{figpage2}}
\end{figure}

Figure \ref{figpage1} shows examples of these modes. As can be seen from
Figure \ref{figpage2} many stars exhibit modes 1-5 in combination. The
$V$$-$$R$ colours show that outburst events result in redder colours. Whilst
modes 1,2 and 5 bring about a simultaneous change in colour, modes 3 and 4
often exhibit a long delay ($\sim$ 1000d) between attaining maximum light and
attaining reddest colour.

The seven year baseline of the MACHO photometry shows the ephemeral nature
of each mode. A particular mode of behavior may switch on or off without
precursor. For this reason we propose that the five modes of variability are
expressions of one underlying mechanism which we discuss below. In a small
number of our sample (324 from 1279 stars) the variability appears
quasi-periodic albeit with variable amplitude.  Amongst quasi-periodic
stars, periods range from 150-1400d for mode 1 with a mean period of 560d to
20-100d for objects of mode 2 (mean P=65d).

%\subsection{Colour-Magnitude Diagram}

\begin{figure}
  \plotone{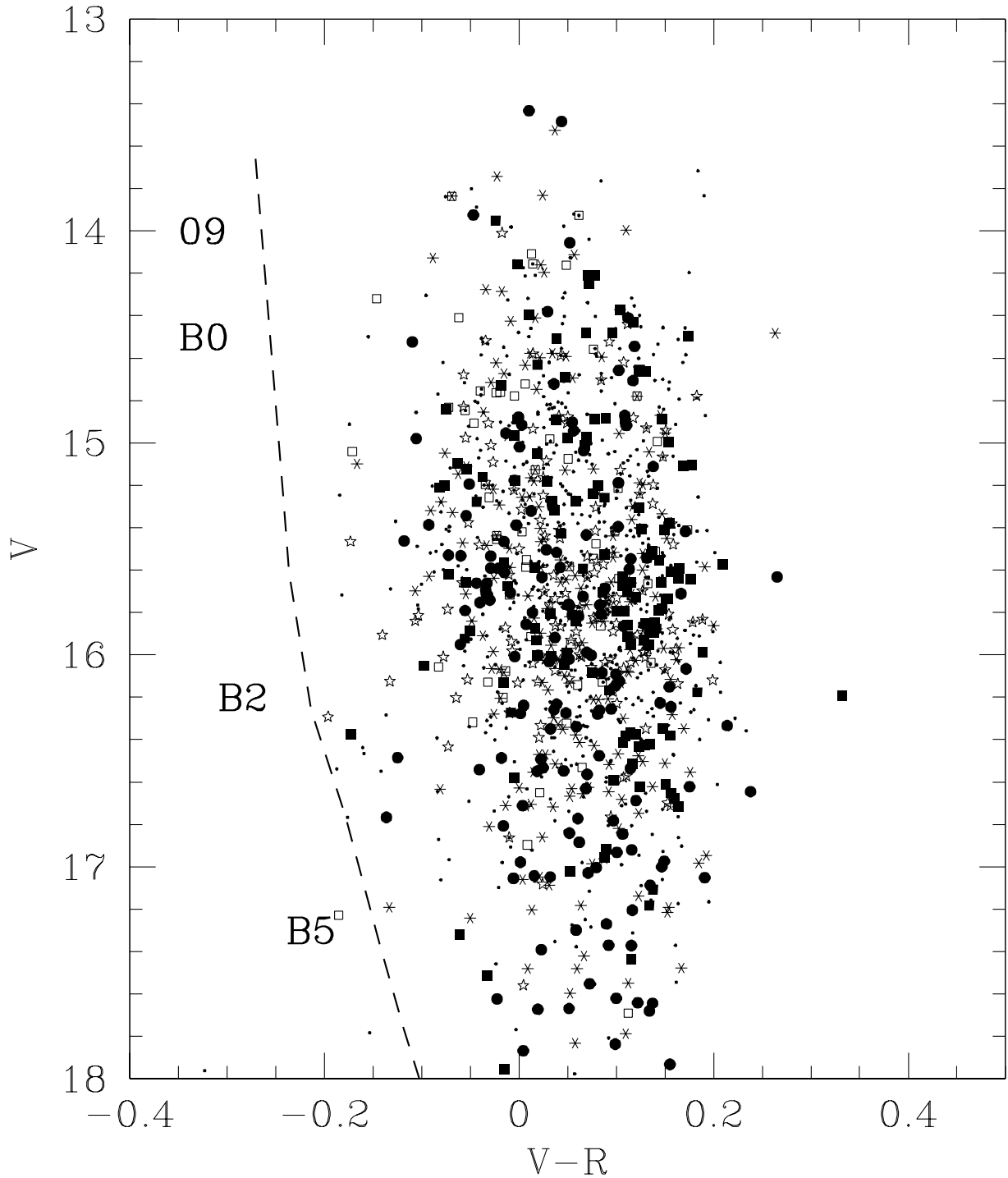}
\caption{The $V$,$V$$-$$R$ colour-magnitude diagram for the blue variables
  described in the present work. Symbols are as follows: mode 1 - solid
  squares, mode 2 - stars, mode 3 - open squares, mode 4 - asterisk and mode
  5 - solid circles. The zero-age main-sequence is shown for comparison
  (Balona \& Shobbrook 1984) \label{figcmd}}
\end{figure}

\begin{figure}
  \plotone{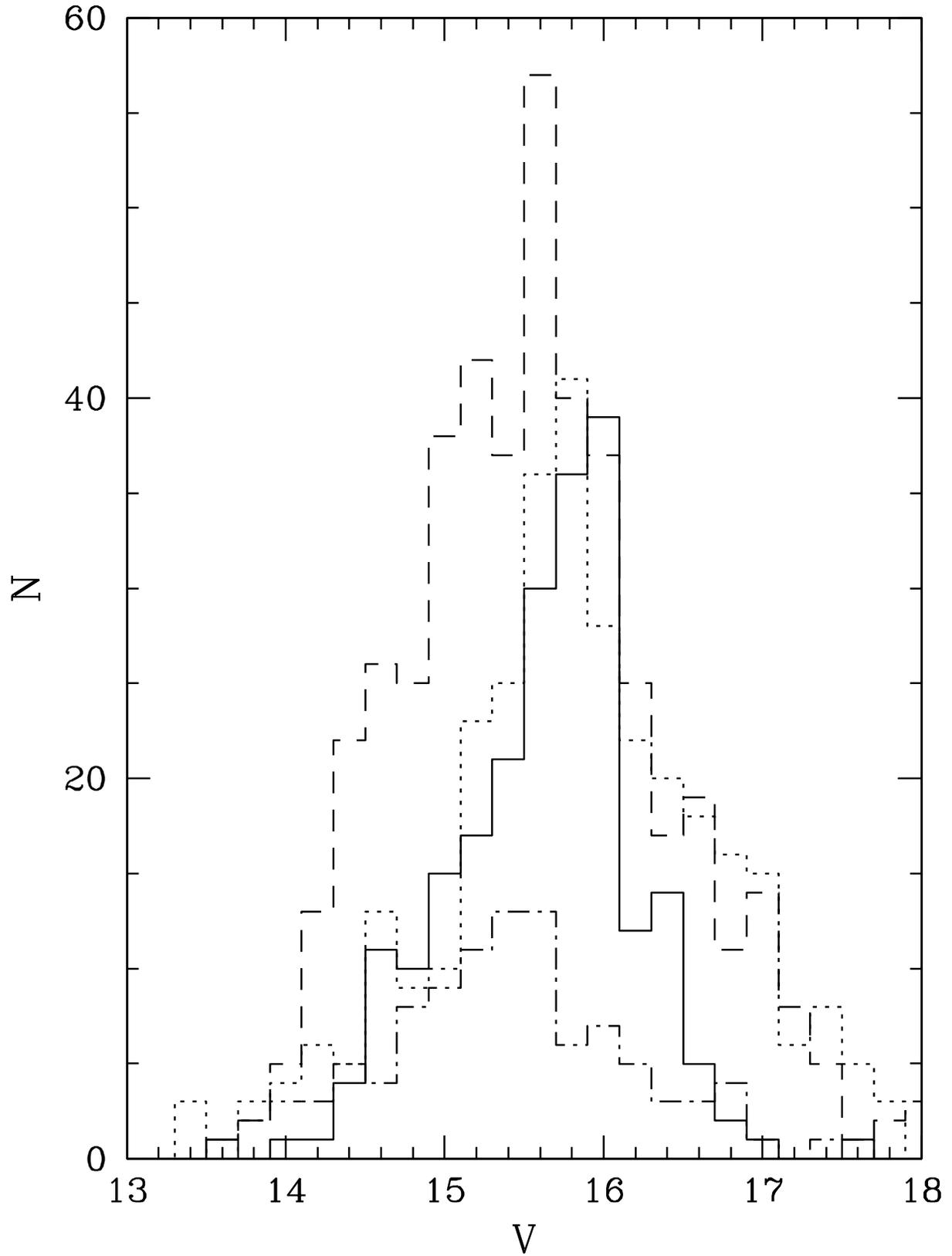}
\caption{Luminosity functions of those stars exhibiting only one mode of
  variability over the duration of our monitoring. Mode 1 is shown dashed,
  mode 2 solid, mode 4 dotted and mode 5 dash-dotted. Mode 3 is not shown as
  it constitutes a very small population.
  \label{figlf}}
\end{figure}

Figure \ref{figcmd} shows the $V$, $V$$-$$R$ colour-magnitude diagram (CMD)
for the sample of blue variables we have isolated. These stars form a broad
band parallel to the non-variable stars on the main-sequence.  This red
offset is of similar magnitude to that observed amongst the Magellanic Cloud
Be star population \citep{kel99}. Shown in large symbols are those objects
which exhibit only one mode in the available data. The five modes inhabit a
similar range in colour. Figure \ref{figlf} presents the luminosity
functions for the subset of single mode stars. At high luminosity, large
outburst behaviour (mode 1) dominates the blue variable population.  The
smaller mode 2 events are significantly skewed to lower luminosities.

%Those objects exhibiting mode 1 behavior (outbursts of $\Delta V >0.2$ mag.
%and durations of $>$100d) present a luminosity function which is
%significantly skewed to higher luminosities when compared to the overall
%distribution (see figure \ref{figdist}). The luminosity functions of other
%modes do not differ significantly.

\section{Spectroscopy}
\label{sect:spec}
Spectroscopy was performed on a subset of 102 blue variables. These
observations were conducted on eight epochs however only 39 stars have two
or more observations. Table \ref{tab:obs} describes our spectroscopic
observations. The sample of objects observed spectroscopically contains
stars from each of the five modes discussed above.

\begin{figure}

\plotone{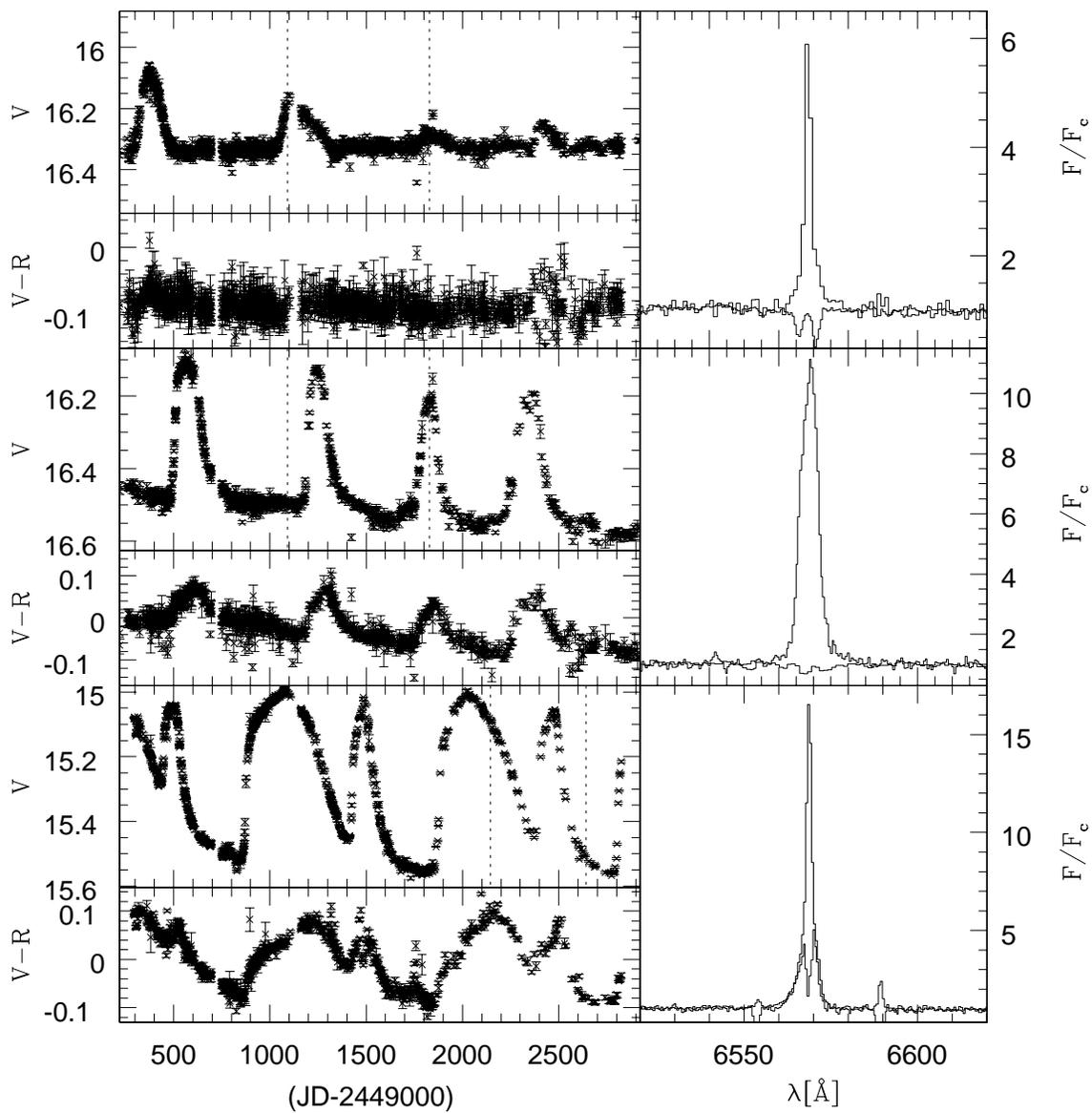}
\caption{Three objects with multiple spectroscopic observations 
  (Top:79.4780.145, Middle:6.6571.71, Bottom:78.6223.41). Note the extreme
  spectra variability between the two epochs indicated by the dotted
  vertical lines on the lightcurve. In each case the epoch of maximum
  emission coincides with that of maximum light.\label{figvar}}
\end{figure}

Amongst the stars in our sample, 91\% show Balmer line emission in at least
one epoch. That is, the majority of objects are Be stars. Given the fact
that such a high proportion of the sample show emission in our limited
number of epochs and the variable nature of the emission-line phenomenon it
is highly likely that all stars in our sample are Be stars. Amongst the 39
stars with multiple spectroscopic observations $\sim50$\% show dramatic
variation in emission equivalent width of H Balmer lines. The most variable
are those which undergo outburst from a quiescent constant baseline. An
example of this extreme variability is shown for three stars in Fig.\
\ref{figvar}.  The H$\alpha$ line profile is seen to vary from absorption to
equivalent widths of over -60\AA. Time of maximum light is associated with
maximum Balmer line emission in each instance.  Those which do not exhibit a
constant baseline are characterised by strong emission (typically $-20$ -
$-40$\AA).

\section{The Nature of Blue Variables}
\label{sect:nature}
The blue variables of the LMC exhibit a variety of lightcurve morphology yet
present similar colour, luminosity distribution and spectral type. We
believe that the lack of differences between objects exhibiting different
modes is due to the fact that the variability has a common underlying
mechanism. The Be phenomenon, and the process underlying it, is the prime
suspect for the production of the variability. A Be star is defined simply
as a B star that has at one time exhibited Balmer emission. We find that
91\% of the blue variable population examined spectroscopically are Be
stars.

An early reference to the unusual population of blue variables in the LMC
was made by \citet{hod61} who discussed a class of variable star that
matches that outlined here, namely, stars to the red of the main-sequence
presenting eruptive outbursts which are redder when brighter. Subsequently,
the EROS microlensing survey have also discussed the blue variable
population in the LMC \citep{bea96,lam99}. These studies highlighted a small
sample of seven objects which satisfy selection criteria similar to those
applied here.  These studies claim however, that the blue variables belong
to a pre-main-sequence (PMS) population analogous to Galactic Herbig Ae/Be
stars.  Study of the EROS SMC data by \citet{bea01} and \citet{dew02}
reveals a further seven similar objects.

The five of the seven EROS LMC objects are present within our catalogue of
blue variables. The counterparts of two objects are not present because they
do not exhibit significant variability in our dataset. Seen in the context
of our eight year baseline rather than the limited 117 days of EROS, these
objects do not show variability which sets them apart from the remainder of
our sample. 

\begin{figure}

\plotone{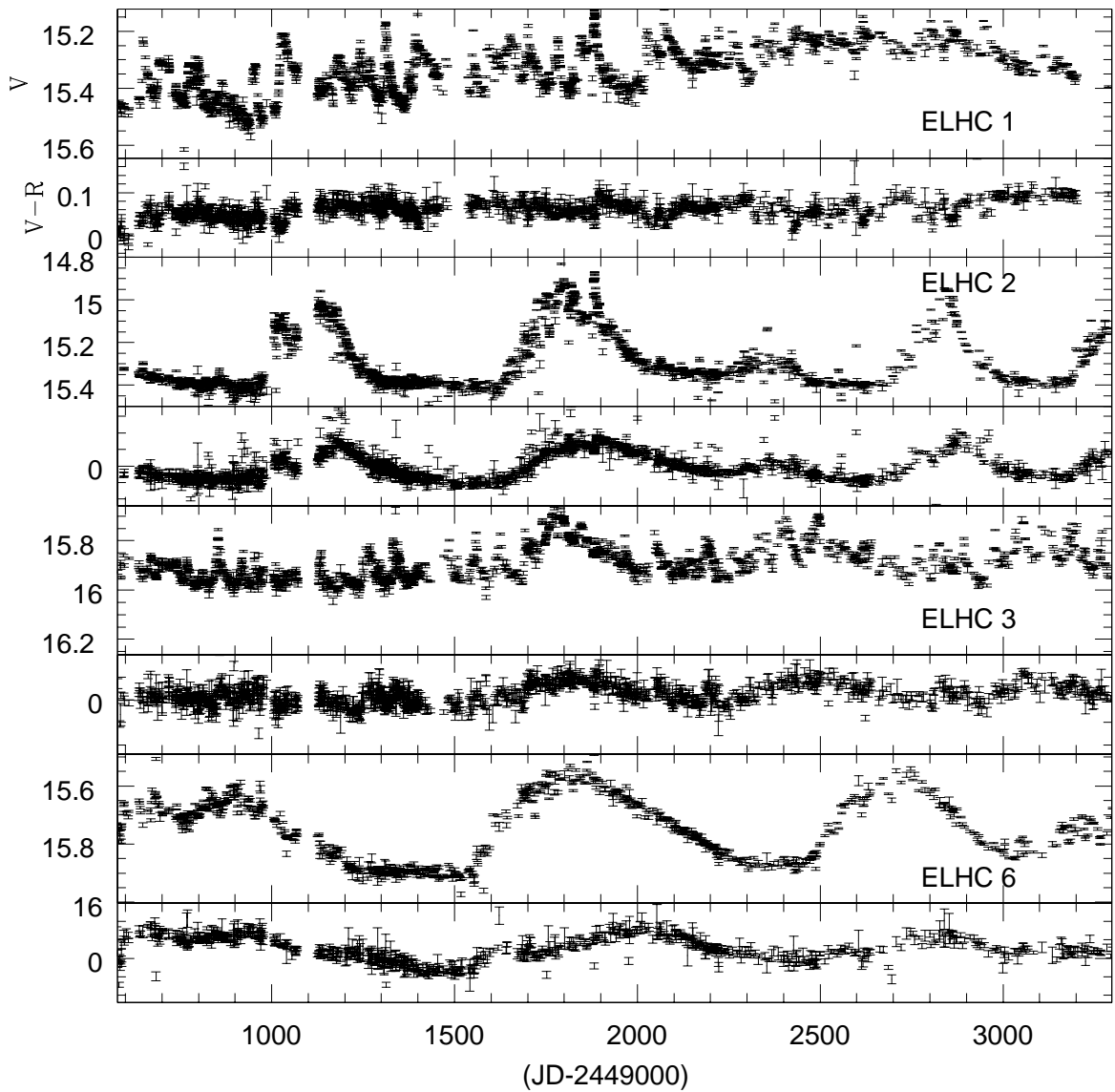}
\caption{$V$ lightcurves and $V$$-$$R$ colourcurves for four 
  pre-main sequence candidates identified by \citet{lam99} (MACHO
  identifications are 78.6101.55(top), 78.6220.76, 79.5863.73, 78.6100.101).
  Our eight year baseline shows that these objects do not presently
  variations of significantly different nature to those presented in Figs.\
  \ref{figpage1} \& \ref{figpage2}.
 \label{fig:EROS}}
\end{figure}

Figure \ref{fig:EROS} presents the lightcurves for ELHC1-3 \& 6 (ELHC5 was
observed in MACHO R band only but is clearly of mode 1). The variability
exhibited is distinct from the frequent, large amplitude (up to 3 mag.)
fading events of short duration (2-50 days) seen in Galactic Herbig Ae/Be
stars \citep{ros01,bib91,the94}. To definitively distinguish between the
possible Be and Herbig Ae/Be nature of these objects will require further
study particularly infra-red spectroscopy.

We propose that the variability of these stars is the result of processes
related to the establishment, maintenance and dissipation of the Be disk.The
emission that characterises Be stars originates in a gaseous circumstellar
quasi-Keplerian disk. No definitive model has been developed which can
sufficiently describe the formation and maintenance of this disk material.
The wind compressed disk model \citep{bjo93} utilises the
combination of fast photospheric rotation with large mass loss rates to
generate a confined disk of material but falls short of generating the
densities observed within Be star disks.  Other models such as the
``one-armed'' density wave model \citep{oka97} seek to explain spectral
line variations observed in Be stars.
 
An insight into the establishment of Be disks is provided by the works of
Rivinius et al.\ (1998; 2001) in their detailed study of the Be star $\mu$
Cen.  These authors find evidence for non-radial pulsation modes within the
star which when in constructive interference provide sufficient mechanical
work to expel material from the stellar surface.  Such material gives rise
to short lived spectral variations which damp as the material melds into the
bulk of the Keplerian disk. The outbursts on $\mu$ Cen have remained of
low intensity and result in only minor modulation of the $V$ luminosity. 

A large proportion of galactic Be stars are known to exhibit photometric and
spectroscopic variability. Many studies exist in the literature which have
sought to characterise the nature of these variations
\citep{fei79,dac82,hub98}. The study by Hubert \& Floquet of Hipparcos
photometry of 289 Be stars shows that amongst early type Be stars
variability greater than 0.$^m$02 (in the broadband Hipparcos photometric
system) is commonplace (98\% amongst B0-B3 stars). Hubert \& Floquet
describe five types of variability which encompass those defined for the
present sample.  Short-term variations (hours to days) are low amplitude (a
few hundredth of a mag.) and are frequently periodic in nature.  Mid-term
variations (days to weeks) are seen to occur as outbursts of several tenths
of a magnitude which are both short-lived ($\sim$50 days: e.g.\ $\omega$
CMa; like mode 2) and long-lived ($\sim$500 days: e.g.\ $\nu$ Cyg; like mode
1) and in the form of fading events. Long-term variations (years) of upto
0.$^m$12 are seen over the 1200d Hipparcos timespan.

The photometric behaviour of Galactic Be stars as described by Hubert \&
Floquet mimics that seen amongst the present sample. In particular the
bumper and flicker outburst modes are of similar duration and amplitude.
Whilst the present sample of blue variables is undoubtably hetrogeneous we
believe that the variability of the majority of objects can be best
understood in connection with the Be phenomenon.

Let us now interpret our observations in terms of the Be phenomenon. Our
spectroscopy indicates that objects exhibiting mode 1 or 2 outbursts from a
flat or quiessent baseline exhibit maximum line emission when in outburst
and an absence of emission when in quiessence. Each outburst event can be
clearly interpreted as a mass-ejection event from the stellar surface.  The
ejection of material into the circumstellar environment creates a rise in
luminosity and redder colour due to the addition of Balmer continuum
emission from the cooler disk component \citep{zor91}. A slow decline
follows the initial outburst as material is removed from the circumstellar
environment due to outflow (see Rivinius et al.\ 2001). In this proposed
scenario, mode 1 and 2 outbursts arise from the same mechanism, however, in
mode 1 more substaintial mass loss occurs. Outbursts may also occur in
systems with established circumstellar material leading to the variety of
lightcurve morphology seen in Fig.\ \ref{figpage2}.  Similarly, fading
events (mode 5) reflect the reduction of circumstellar material.
Unfortunately we do not have multi-epoch spectroscopic observations over a
fading event to confirm this.

The blue variables we have presented in this work afford an excellent probe
into the mechanisms of establishment and maintainence of the Be phenomenon.
Further study of the combined photometric and spectroscopic variations may
infer the mechanisms responsible for mass ejection and establishment of the
circumstellar disk.

\section{Summary} 

We have presented a summary of the properties of 1279 blue variables within
the Large Magellanic Cloud. Variations of up to $\Delta V \sim$0.$^m$4 are
observed in the form of rapid ($\sim$50d) or prolonged (100-500d) eruptive
events as well as fading events and long term (years) trends. The $V$$-$$R$
colours show a characteristic ``redder when brighter'' behaviour.

A subset of 102 stars were examined spectroscopically, in a limited number
of instances over multiple epochs.  Many objects show spectacular
photometric and spectroscopic variability. The majority (91\%) of the
spectroscopic sample were found to exhibit Balmer emission lines in at least
one epoch (i.e.\ Be stars).

This strongly suggests that the mechanism giving rise to the variability
seen with in the sample of LMC stars is related to the Be phenomenon.
Firstly, the photometric variability is similar to that observed in Galactic
Be stars. Secondly, emission strength is seen to coincide with redder
$V$$-$$R$ colour and maximum $V$ luminosity as expected from known
properties of Galactic Be stars.

The outbursts observed in the blue variable population can therefore be
understood as mass eject events from the central star leading to the
establishment and maintenance of the Be disk. Through further study of this
population could yield important insight into the physical process of mass
loss and the confinement of material in the vicinity of star which underlies
the Be star phenomenon. The role of metallicity in the Be phenomenon will be
examined in a future paper where we plan to investigate properties of the
blue variable population of the Small Magellanic Cloud.

\begin{acknowledgements} 
  
  This work was performed under the auspices of the U.~S.\ Department of
  Energy, National Nuclear Security Administration by the University of
  California, Lawrence Livermore National Laboratory under contract
  W7405-Emg-48.
\end{acknowledgements}

\clearpage



%\begin{figure}
%\plotone{dist.eps}
%\caption{The $V$ luminosity function for the sample of blue variables (total 
%sample - solid line; mode 1 sample - dashed line). \label{figdist}}
%\end{figure}

\clearpage
 
\begin{table}
\begin{center}
\caption{Details of our spectroscopic observations\label{tab:obs}}
\begin{tabular}{ccccc}
\tableline\tableline
date & MJD & Instrument & $\lambda$(\AA) & Resolution $\Delta \lambda$(\AA)\\
\tableline
30 Dec 1994 & 49717 & AAT/RGO & 6290-6840 & 1.0\\
2  Jan 1997 & 50451 & CTIO 60'' & 3640-7330 & 4.4\\
26 Nov 1998 & 51144 & SSO 2.3m/DBS & 3600-4580; 6095-7050&  1.1; 1.1\\ 
28 Mar 1999 & 51266 & SSO 2.3m/DBS & 3315-5500; 5500-10200& 4.4; 8.0\\ 
22 Sep 1999 & 51444 & SSO 2.3m/DBS & 3250-6410; 6400-10400& 4.4; 8.0\\ 
12 Dec 1999 & 51525 & SSO 2.3m/DBS & 3600-4600; 6100-7050&  1.1; 1.1\\ 
2  Dec 2001 & 52246 & SSO 2.3m/DBS & 3610-4590; 6090-7045&  1.1; 1.1\\ 
4  Mar 2002 & 52310 & SSO 2.3m/DBS & 3610-4590; 6090-7045&  1.1; 1.1\\ 
\tableline
\end{tabular}
\end{center}
\end{table}

\begin{table}
\begin{center}
\caption{Positions and mean photometry for the sample of 1279 blue variables. 
  The full table is available in electronic form from ADS. The column ``mode''
  describes the most prominent of the five modes outlined in section 3. In
  some cases where a number of modes are apparent these are also given.  The
  full table is available from ADS.\label{tab:data}}
\begin{tabular}{cccccc}
\tableline\tableline
MACHO id & $\alpha$(J2000) & $\delta$(J2000) & $V$ & $V-R$ & mode \\
\tableline
1.3321.7    & 05:01:13.700 & -69:20:04.24 & 14.756 & 14.671 & 1\\
1.3442.19   & 05:01:27.791 & -69:19:51.16 & 14.747 & 14.729 & 5\\
1.3446.1061 & 05:01:35.476 & -69:04:28.83 & 16.564 & 16.494 & 4\\
1.3563.47   & 05:02:32.764 & -69:21:18.40 & 15.756 & 15.676 & 24\\
1.3564.154  & 05:02:31.128 & -69:14:36.92 & 17.182 & 17.048 & 1\\
\ldots&        \ldots      & \ldots       & \ldots  & \ldots  & \ldots  \\
\tableline
\end{tabular}
\end{center}
\end{table}


\begin{thebibliography}{}
%\bibitem[]{} 
\bibitem[Alcock et al.(1995)]{alc95} Alcock, C.~et al.\ 1995, \aj, 109, 1653
\bibitem[Alcock et al.(1999)]{alc99} Alcock, C.~et al.\ 1999, \pasp, 111, 1539
\bibitem[Balona \& Shobbrook(1984)]{bal84} Balona, L.A., \& Shobbrook, R.R.\ 1984, MNRAS, 211, 375
\bibitem[Beaulieu et al.(1996)]{bea96} Beaulieu, J.~P.~et al.\ 1996, Science, 272, 995
\bibitem[Beaulieu et al.(2001)]{bea01} Beaulieu, J.-P.~et al.\ 2001, \aap, 380, 168
\bibitem[Bibo \& The(1991)]{bib91} Bibo, E.A., \& The, P.S.\ 1991, \aaps, 89, 319
\bibitem[Bjorkman \& Cassinelli(1993)]{bjo93} Bjorkman, J.~E.~\& Cassinelli, J.~P.\ 1993, \apj, 409, 429
  
\bibitem[Cook et al.(1995)]{coo95} Cook, K.~H.~et al.\ 1995, in Astrophysical
  Applications of Stellar Pulsation, ASP Conf.~Ser.~83, ed.\ R.~S.~Stobie \&
  P.~A.~Whitelock (San Francisco:ASP), 221

\bibitem[de Wit, Beaulieu \& Lamers(2002)]{dew02}de Wit, W.J.M., Beaulieu, J-P. \& Lamers, H.\ A\&A, submitted
  
\bibitem[Dachs(1982)]{dac82}Dachs, J.\ 1982, in IAU Symp.\ 98, Be stars, ed. M.
  Jaschek \& H.-G. Groth (Dordrecht: Reidel), 19

\bibitem[Feinstein \& Marraco(1979)]{fei79}Feinstein, A., \& Marraco, H.G.\ 1979, AJ, 84, 1713
\bibitem[Hodge(1961)]{hod61}Hodge, P.W.\ 1961, Obs 81, 31
\bibitem[Hubert \& Floquet(1998)]{hub98}Hubert, A.M., \& Floquet, M.\ A\&A, 335, 565
\bibitem[Keller, Wood \& Bessell(1999)]{kel99} Keller, S.~C., Wood, P.~R., \& Bessell, M.~S.\ 1999,
  \aaps, 134, 489 
\bibitem[Lamers, Beaulieu \& de Wit(1999)]{lam99} Lamers, H.~J.~G.~L.~M., Beaulieu, J.~P., \& de Wit, W.~J.\ 1999, \aap, 341, 827 
\bibitem[Okazaki(1997)]{oka97} Okazaki, A.~T.\ 1997, \aap, 318, 548
\bibitem[Rivinius et al.(1998)]{riv98} Rivinius, T., Baade, D., {\v S}tefl, S., Stahl, O., Wolf, B., \& Kaufer, A.\ 1998, \aap, 333, 125
\bibitem[Rivinius et al.(2001)]{riv01} Rivinius, T., Baade, D., {\v S}tefl, S., \& Maintz, M.\ 2001, \aap, 379, 257
\bibitem[Rostopchina \& Grinin(2001)]{ros01} Rostopchina, 
A.~N.~\& Grinin, V.~P.\ 2001, Astronomy Reports, 45, 51 
\bibitem[The(1994)]{the94} The, P.~S.\ 1994, in ASP Conf.~Ser.~62: The
  Nature and Evolutionary Status of Herbig Ae/Be Stars, ed. P.~S.~The,
  M.~R.~Perez, and E.~P.~J.~Van den Heuvel (San Francisco: ASP), 23
\bibitem[Zorec \& Briot(1991)]{zor91} Zorec, J.~\& Briot, D.\ 1991, \aap,
  245, 150
  
\end{thebibliography}
\end{document}